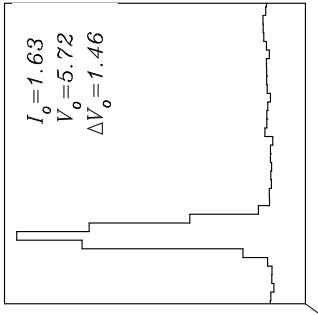
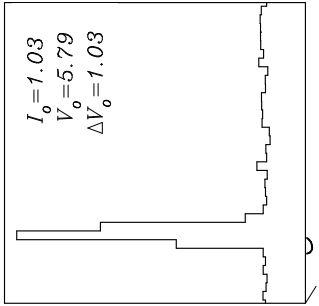
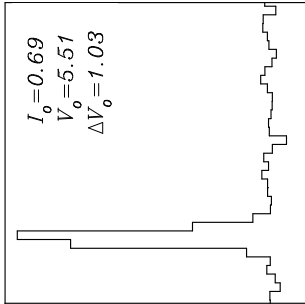
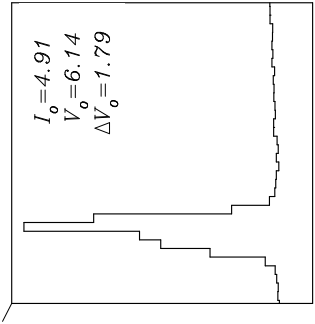
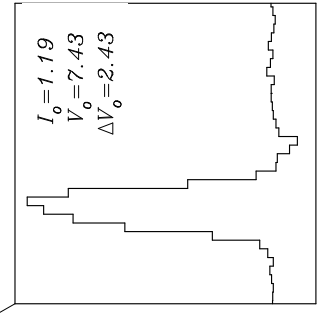
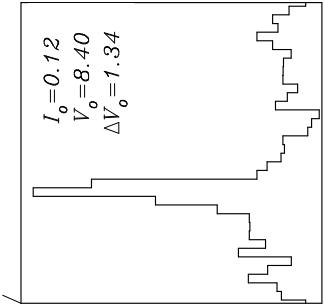
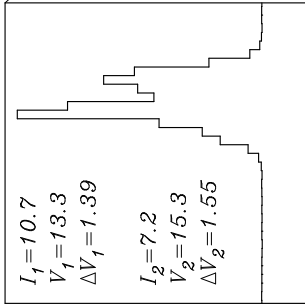
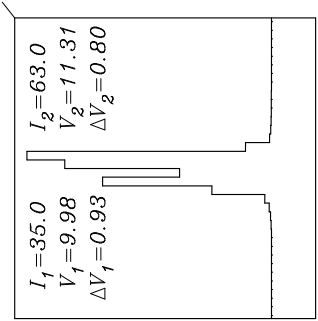
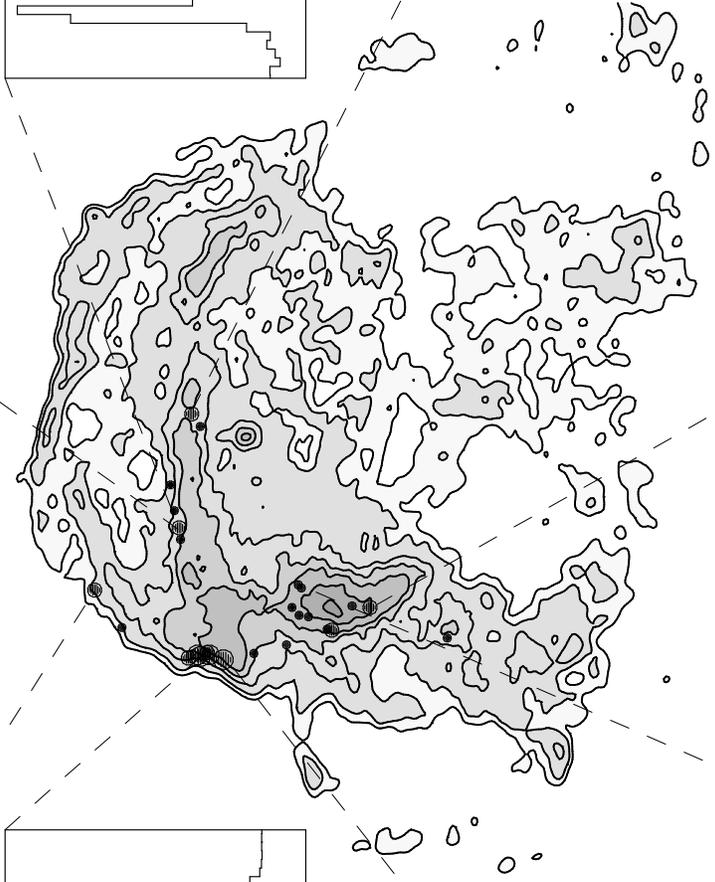

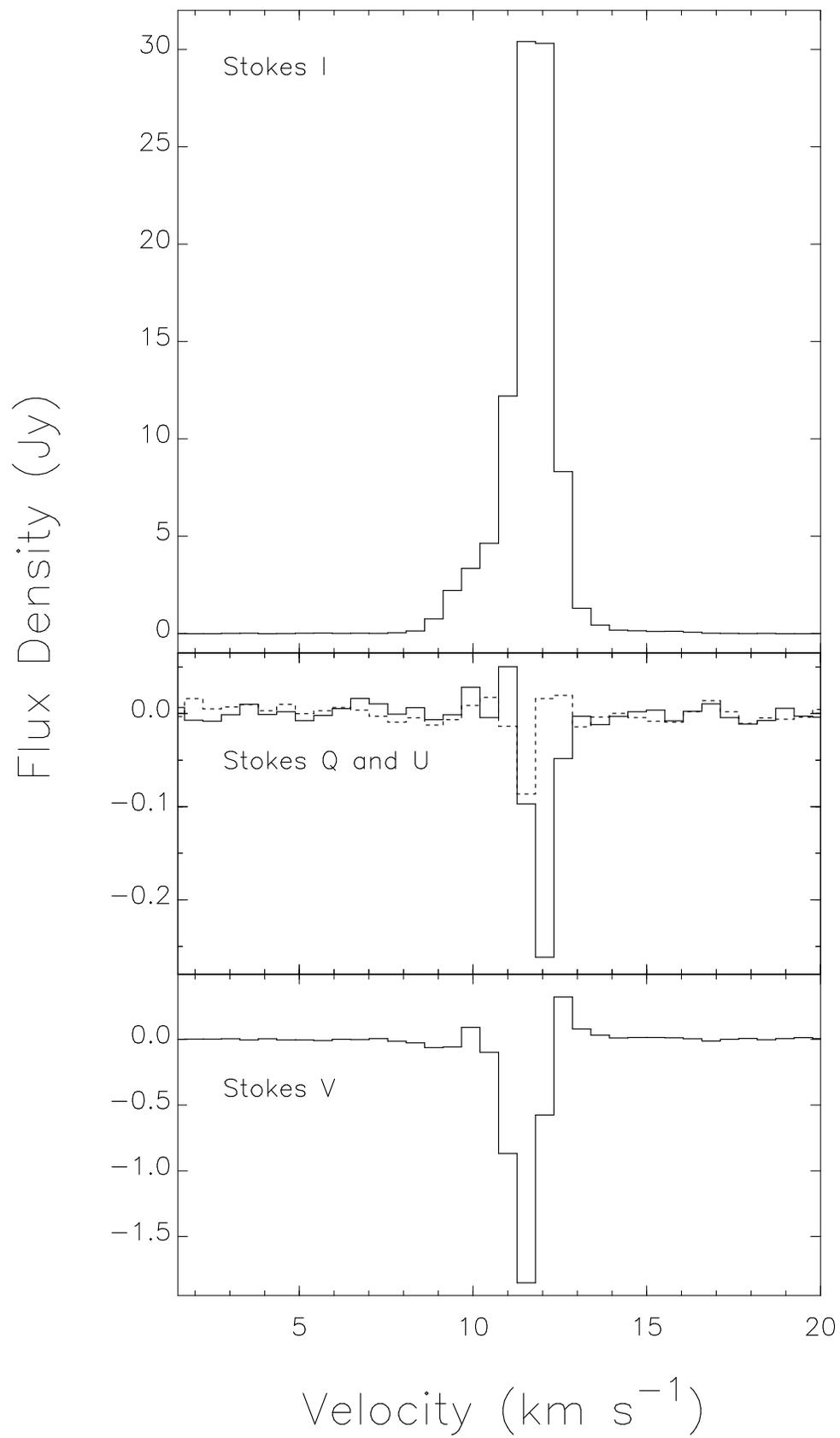

# Shock Excited Maser Emission from the Supernova Remnant W 28

D. A. Frail  *NRAO/AOC, Socorro, New Mexico, U.S.A.*

W. M. Goss  *NRAO/AOC, Socorro, New Mexico, U.S.A.*

V. I. Slysh  *Astro-Space Center, Moscow, Russia*

**Summary.** We present VLA observations in the hydroxyl line OH(1720 MHz) towards the supernova remnant W 28. These high resolution images show a more extensive distribution of 1720 MHz emission than had previously been suspected from earlier single-dish observations. A total of 26 distinct maser spots were detected which lie on the edge of the radio continuum emission from W 28 and are unresolved with our $12''$ resolution. Significant detection of linear and circular polarization in the maser spots is also made. The location of the maser spots, together with the physical conditions in the SNR and the adjacent molecular cloud, argue for a shock excitation mechanism. The possibilities of using the OH(1720 MHz) as a powerful diagnostic of the interaction of supernova remnants with molecular clouds is discussed.

# 1. Introduction

In the life cycle of the stars and gas in our Galaxy it is the interaction of supernova remnants (SNRs) with molecular clouds (MCs) that joins the two endpoints of this evolutionary process. In recent years the emphasis in the study of such interactions has understandably shifted away from centimeter wavelengths to the millimeter and sub-millimeter wavelengths, where important transitions from shocked molecular species are in abundance (Dishoeck, Jansen & Phillips 1993). In this paper we wish to call attention to an overlooked but potentially powerful probe of SNR/MC interactions: the 1720 MHz line of the hydroxyl radical (OH).

While galactic maser emission has been well-studied from OH both in late-type stars and in HII regions, a possible counterpart near SNRs has been largely forgotten since its discovery (Goss 1968). Goss & Robinson (1968), Turner (1969), and Robinson, Goss & Manchester (1970) detected strong OH line emission at 1720 MHz and broad-line absorption at 1612, 1665 and 1667 MHz in two SNRs: W 28 and W 44. DeNoyer (1979) also observed similar lines toward the SNR IC 443. These OH lines are distinguished from their HII region counterparts in having emission *only* at 1720 MHz, relatively low degrees of linear and circular polarization (5-10%), and somewhat wider line widths (1-2 km s$^{-1}$).

It is noteworthy that all three of these SNRs with strong 1720 MHz OH emission have also been shown to be interacting with dense molecular clouds (W 28: Wootten 1981, Pastchenko & Slysh 1974; W 44: Wootten 1977; IC 443: DeNoyer 1979). Pastchenko & Slysh (1974), by mapping 1667 MHz main line OH absorption, have shown that the molecular cloud is adjacent to the supernova shell and has a crescent-shaped appearance with the radius of curvature pointing to the center of the supernova remnant. The position of the 1720 MHz maser was found to be in the middle of the molecular cloud. They suggested that the cloud was shaped by an interaction with the supernova envelope and that the location of the 1720 MHz maser in the dense molecular cloud in the immediate vicinity of the expanding supernova shell was indicative of the maser emission being pumped by the shock.

From CO mapping Wootten (1981) points out additional evidence to support an interaction: a) the CO line width broadens significantly where the cloud meets W 28, (b) the cloud is warmer than "typical" CO clouds in the Galaxy, and (c) emission from $^{13}$CO and HCO$^+$ is strongest near W 28. Thus it appears that the 1720 MHz OH emission arises as a result of this interaction. Since previous single-dish observations did not have enough angular resolution to be able to accurately locate the position of the masers with respect to W 28 and the molecular cloud, we undertook the present study to better understand the relationship between the 1720 MHz OH emission and the SNR W 28.

## 2. Observations

The observations were made at the VLA [1] on 1993 June 14 in its C-array config-

---

**Note 1:** The Very Large Array (VLA) is operated by the National Radio Astronomy Observatory under cooperative agreement with the National Science Foundation.

---

uration. A bandwidth of 195 kHz and 64 frequency channels were recorded in the parallel and cross hands of circular polarization for a resolution of 3.05 kHz per channel (or 0.53 km s$^{-1}$), giving a total velocity coverage of 34 km s$^{-1}$. The center of the band was at +11 km s$^{-1}$ (LSR) and the pointing center was chosen to be at the peak of the maser emission, as measured by Robinson, Goss & Manchester (1970). The data were calibrated within the AIPS package following standard practice.

Three-dimensional image cubes were made in each of the 4 Stokes parameters, *I*, *Q*, *U* and *V*. The channel containing the brightest maser emission in *I*-data was used to self-calibrate the entire dataset. Solutions obtained from this process were applied to all channels and all Stokes parameters. These images were then CLEAN'ed reaching a final rms noise level of 7 mJy beam$^{-1}$ for a beam size of 12″. The peak flux density in the I-images was 63 Jy. Once individual maser components were identified from the total intensity image cube, the spectra were extracted from the cubes at these positions (for all Stokes parameters), and Gaussian fits were made to the individual line components.

## 3. Results

A total of 26 distinct maser "spots" were identified in the data, all of which appear to be unresolved with our $12''$ beam. Their positions and relative brightnesses are shown in Figure 1. There is little doubt that we are observing maser emission. The brightness temperature limits ($T_b > 4 \times 10^5$ K) derived from the peak flux density, together with the small line widths ($\Delta V \simeq 1$ km s$^{-1}$) point to a non-thermal origin for this emission. Masers were found at the two locations noted by Robinson *et al.* (1970), but in addition we detect masers to the north and west, outside of their $12'$ beam and beyond the region mapped in the $^{12}$CO and $^{13}$CO lines by Wootten (1981). The peak flux densities of the masers vary from 0.1 Jy to 63 Jy. While the maser emission appears to be clustered along the eastern edge of W 28, this may only be an observing artifact, a consequence of our pointing center and our finite field of view (radius $15'$).

The maser spots are located along the *edge* of the continuum radio emission from W 28. There is a strong preference for locations at which there is a gradient in the radio emission, and in fact, the strongest maser emission arises where the gradient is steepest. All of the 1720 MHz emission is coincident with W 28, despite the fact that there are several prominent HII regions in the vicinity, including the well-studied compact HII region G6.6$-$0.1 (Andrews, Basart & Lamb 1985).

Total intensity spectra are shown for a representative subset of the profiles in Figure 1. While our observations were sensitive to lines with a range of velocities between $+26$ km s$^{-1}$ and $-4$ km s$^{-1}$, all of the maser emission was found at velocities between $+5$ km s$^{-1}$ and $+15$ km s$^{-1}$. This same velocity range was observed by Hardebeck (1971) with the Owens Valley interferometer. The great majority of spectra have single gaussian profiles with velocity centroids between $+6$ km s$^{-1}$ and $+8$ km s$^{-1}$, close to the velocity of the molecular cloud at $+7$ km s$^{-1}$ (Wootten 1981, Pastchenko & Slysh 1974). The large positive velocities ($> +8$ km s$^{-1}$) are from masers located along the eastern edge of W 28 which are seemingly in direct contact with the molecular cloud mapped by Wootten (1981). Double-peaked profiles are rare and are only found at the brightest cluster of maser spots. Two of the spectra (the two bottom spectra in Figure 1) show some evidence for absorption. Both masers are located near the continuum peak of W 28. The absorption for the higher quality profile is displaced by $+4.1$ km s$^{-1}$ from its emission peak.

**Figure 1**

Linear and circular polarized flux was detected for all of the bright maser spots near the pointing center at the level of 1 to 10%. This is similar to the degree of polarization seen by previous investigators (e.g. Goss & Robinson 1968). A sample profile is shown in Figure 2.

**Figure 2**

## 4. Discussion

The location of the 26 maser spots along the edge of the radio continuum of W 28, and the close match between their velocities and that of an adjacent molecular cloud, suggests that it is this interaction which is responsible for this unique OH 1720 MHz emission. Litvak (1974) suggested a radiative pump to invert the 1720 MHz line but in the far-IR at least there are few sources to provide the excitation (Odenwald *et al.* 1984). Moreover, Elitzur (1976) showed that radiative pumps can produce only weak 1720 MHz OH emission, while collisions with $H_2$ molecules with kinetic temperatures $T_K$ below 200 K can strongly invert the 1720 MHz line. The inversion functions efficiently within a range of temperatures between 25-200 K and densities between $10^3$-$10^5$ cm$^{-3}$. Above 200 K the 1612 MHz transition begins to dominate and for $H_2$ densities above $10^5$ cm$^{-3}$ collisions depopulate the excited states quenching the maser process.

The physical conditions in W 28 and the molecular cloud match the requirements of Elitzur's maser model well. The pre-shock temperature and average density of the molecular cloud is $T_o$=15 K and $n_o$=2.5×$10^4$ cm$^{-3}$ (Wootten 1981; DeNoyer 1983). Fabry-Perot observations of H$\alpha$ filaments by Lozinskaya (1974) measure a mean expansion velocity $\overline{V}_s$ for W 28 of 40-50 km s$^{-1}$, but some filaments have $V_s$ as high as 80 km s$^{-1}$. Optical line ratios (Bohigas *et al.* 1983; Long *et al.* 1991) give values for shock velocities that are typically twice the $\overline{V}_s$ from the H$\alpha$ measurements. For these values of $V_s$ and $n_o$ a J-type shock is likely to be present (Draine & McKee 1993) and significant amounts of OH form in the required temperature range of 25-200 K at a distance of 20-100 AU behind this dissociative shock (Hollenbach & McKee 1989, Neufeld and Dalgarno 1989).

We can also estimate the size of the masing region from the existing data. Estimates of the post-shock column density of OH in W 28 range from $10^{16}$ cm$^{-2}$ to $10^{17}$ cm$^{-2}$. Following Pastchenko & Slysh (1974) we adopt 1.4×$10^{16}$ cm$^{-2}$ ($T_s$=20 K). At a distance of 2 kpc the 5′ × 15′ OH cloud has dimensions of 3 pc × 9 pc and a velocity dispersion of $\Delta V$=16 km s$^{-1}$. Assuming a pathlength through the cloud of r=3 pc we derive $n_{\rm OH}/(\Delta {\rm V}/r) = 2.8 \times 10^{-4}$ cm$^{-3}$/ km s$^{-1}$ pc$^{-1}$. For this value of $n_{\rm OH}/(\Delta {\rm V}/r)$ and T$_K >$ 70 K, Elitzur's (1976) collisional pump model predicts an optical depth $\tau_{1720} \simeq -20$, which amplifies the background continuum emission from W 28 (T$_c \simeq$ 10 K) to T$_b$=4$\times$10$^9$ K (T$_b$ =T$_c$ e$^{20}$). At the peak flux density of the maser spots which we observe (63 Jy), this translates into a spot size of 20 AU (or 10 mas). This predicted size is comparable to the expected thickness of the OH zone described above. This prediction is easily tested by high resolution radio observations.

The post-shock magnetic field can be constrained by these observations. One possible interpretation for the shape of the Stokes V-profiles (Figure 2) is Zeeman splitting of the circular hands of polarization in the presence of a magnetic field. For the OH molecule at 1720 MHz the magnitude of this effect is 1.31 kHz mG$^{-1}$ (Heiles *et al.* 1993). The observed splitting of 6 kHz corresponds to a magnetic field strength of about 5 mG. However, without adequate spectral and angular resolution, the measurement of the strength of the field is problematic (Reid and Moran 1981) but the direction of the field can be inferred from the Stokes V profiles (e.g. Roberts *et al.* 1993). Only the brightest masers, within 3$'$ of the pointing center, have reliable Stokes V profiles. The line-of-sight direction of the field in this region is consistently positive (i.e. points away from the observer). With higher resolution data, combined with the perpendicular component of the field from radio synchrotron measurements (e.g Milne, Caswell & Haynes 1992), a full determination of the field strength and direction will be possible at select points along W 28.

## 5. Conclusions

Since its discovery in the late 1960's, maser emission from the 1720 MHz transition of the OH molecule has not received the full observational attention that it warrants. Our new, high-resolution observations demonstrate that extensive maser activity is taking place along the eastern edge of W 28. The location of the maser spots, together with the physical conditions in the SNR and the adjacent molecular cloud, argue for a different inversion mechanism that is commonly assumed for HII region masers and stellar masers. Our observations support the collisional model of Elitzur (1976) and we suggest that under certain conditions a shock, propagating in a molecular cloud, can collisionally excite OH behind the shock, preferentially populating the 1720 MHz inversion.

In addition to mapping out the full extent of the 1720 MHz maser emission in W 28, future high resolution observations should image the SNRs W 44 and IC 443. A search for 1720 MHz emission towards other SNRs that are known to be interacting with molecular clouds (e.g. Huang & Thaddeus 1986) is bound to be fruitful. For example, OH emission at 1720 MHz can be found at the location of 8 SNRs in the OH survey of Turner (1979), Green (1989) lists several SNR candidates with evidence of shocked OH gas in absorption, and Koo & Heiles (1991) detect high velocity hydrogen associated with SNRs. VLBI observations offer the promise of determining the strength and geometry of the magnetic field in the post-shock gas as well as being able to determine the sizes and proper motions of the maser spots. Early indications are that we have a powerful new probe to study the interaction of supernova remnants with molecular clouds.

**Figure Captions**

**Figure 1.** The distribution of the 26 OH maser spots across the supernova remnant W 28. The radio continuum image is from a 327 MHz project by Frail, Kassim & Weiler (1993). The maser spots are indicated by the black circles and are shown in three sizes, depending upon their peak intensity: small circles (I< 1 Jy), medium circles (I< 10 Jy), and large circles (I> 10 Jy). Also shown are eight spectra, taken as being representative of the total. For each of these spectra are shown the results of a Gaussian fit to the spectral shape for which we derive the peak intensity $I$ in Jy, the central velocity $V$ in km s$^{-1}$ and the line width $\Delta V$ in km s$^{-1}$. The velocity scale increases left to right from 1.5 km s$^{-1}$ to 20 km s$^{-1}$. The east-west angular diameter of W 28 is approximately 40$'$.

**Figure 2.** A sample profile of a maser spot towards the SNR W 28 at $\alpha(1950)=17^h$ $57^m$ $48.91^s$, $\delta(1950) = -23°$ $18'$ $0.0''$. All four Stokes parameters are shown. The linear Stokes parameters in the center panel are distinguished by the line type: a solid line for $Q$ and a dashed line for $U$.